\documentclass{elsart}               

\usepackage{graphicx}          
\usepackage{natbib}                               

\begin{document}

\begin{frontmatter}

\title{The final phase of inspiral of neutron stars: realistic equations of state} 


\author[Alic,Ziel,Paryz,Waw]{D. Gondek-Rosi\'nska}\ead{Dorota.Gondek@obspm.fr},    %
\author[Waw]{M. Bejger},
\author[OA,Waw]{T. Bulik},
\author[Paryz]{E. Gourgoulhon},
\author[Waw,Paryz]{Pawe\l \ Haensel},
\author[Paryz]{F. Limousin}
\author[USA]{K. Taniguchi}
\author[Waw]{L. Zdunik}

\address[Alic]{Departament de Fisica Aplicada, Universitat d'Alacant, Apartat de correus 99, 03080 Alacant, Spain}
\address[Ziel]{Institute of Astronomy, University of Zielona G\'ora, Lubuska 2, 65-265, Zielona G\'ora, Poland}

\address[Paryz]{LUTH, Observatoire de Paris, Universite Paris 7, Place Jules Janssen, 92195 Meudon Cedex, France}  
\address[Waw]{CAMK, Bartycka 18, 00716, Warsaw, Poland}      

\address[OA]{Astronomical Observatory, Warsaw University, Al. Ujazdowskie 4, 00476 Warsaw, Poland}       
\address[USA]{Department of Physics,
University of Illinois at Urbana-Champaign,
Urbana, Illinois 61801, USA}

\begin{abstract}                          

Coalescing compact star binaries are expected to be among the
strongest sources of gravitational radiation to be seen by laser
interferometers. We present calculations of the final phase of
inspiral of equal mass irrotational neutron star binaries and strange
quark star binaries.  Six types of equations of state at zero
temperature are used - three realistic nuclear equations of state of
various softness and three different MIT bag models of strange quark
matter. We study the precoalescing stage within the
Isenberg-Wilson-Mathews approximation of general relativity using a
multidomain spectral method. The gravitational-radiation driven
evolution of the binary system is approximated by a sequence of
quasi-equilibrium configurations at fixed baryon number and decreasing
separation. We find that the innermost stable circular orbit (ISCO) is
given by an orbital instability for binary strange quark stars and by
the mass-shedding limit for neutron star binaries.  The gravitational
wave frequency at the ISCO, which marks the end of the inspiral phase,
is found to be $\sim 1100-1460 $~Hz for two $1.35\, M_\odot$
irrotational strange stars described by the MIT bag model and between
$ 800 ~{\rm Hz}$ and $1230~{\rm Hz}$ for neutron stars.
\end{abstract}

\end{frontmatter}

\section{Introduction}
Coalescing neutron star binaries are considered among the strongest
and most likely sources of gravitational waves to be seen by
VIRGO/LIGO interferometers \cite{Burgay04, Kalogera04, Belczynski02}.
Due to the emission of gravitational radiation, binary compact stars
decrease their orbital separation and finally merge. We separate
the evolution of
a binary system into three phases : {\it point-like
inspiral} where orbital separation is much larger than the neutron
star radius, {\it hydrostationary inspiral} where orbital separation is
just a few times larger than the radius of the neutron star so that
hydrodynamics play an important role, and {\it merger} in which the
two stars coalesce dynamically.  Gravitational waves emitted during
last orbits of inspiral phase and the merger phase could yield
important information about the equation of state (EOS) of dense
matter \cite{Faber02, TanigG03, Oechslin04, Bejger04, LimouGG05}.  Up
to now, all relativistic calculations (except those of
\cite{Oechslin04,Bejger04,LimouGG05}) of the last orbits of inspiral
phase have been done for the simplified EOS of dense matter - the
polytropic EOS.  In the paper we present results of our studies on
late stage of inspiral of binary systems containing equal mass compact
stars described by different realistic EOS of dense matter.   The
calculations are performed in the framework of {\em
Isenberg-Wilson-Mathews} approximation to general relativity (see
Ref. \cite{BaumgS03} for a review). We consider binary systems
consisting of two identical stars. We choose the gravitational mass of
each star to be $1.35 \, M_\odot$ at infinite separation in order to
be consistent with recent population synthesis calculations
\cite{BulikGB04,GondeBB05} and with the current set of well-measured neutron
star masses in relativistic binary radio pulsars
\cite{Lorim01,Burga03}.  We assume that the velocity flow in the
stellar interiors is irrotational (the fluid has zero vorticity in the
inertial frame) since the viscosity of neutron star matter (or strange
star matter) is far too low to ensure synchronization during the late
stage of the inspiral \cite{BildsC92,Kocha92}.

\section{Equations of state and stellar models}
In Fig. 1 we show gravitational mass versus stellar radius for sequences
of static compact stars.  We limit ourselves to neutron stars
consisting of nucleons and hyperons and strange quark stars described
by the MIT bag model.  Depending on the EOS we obtain the radius of a
1.35 $M_\odot$ star in the range 10-14 km.  We perform calculations
for three nuclear EOS of dense matter based on modern
many-body calculations.  The differences in the M-R relation for
stellar models of neutron stars (solid lines) shown in Fig. 1 reflect
the uncertainties in the existing theories of the interactions in
nuclear matter. We consider one soft ({\bf BPAL12}, \cite{Bombaci95})
and one stiff ({\bf AkmalPR}, \cite{APR98}) EOS of matter composed of
nucleons, electrons and muons. We considered also one EOS in which
hyperons are present at high densities ({\bf GlendNH3},
\cite{Glend1985}). The neutron star crust is described by means of a
realistic EOS obtained in the many-body calculations (see
\cite{Bejger04} for details). We note that the equations of state 
of nuclear matter can be considered static, as the timescales
for bulk processes such as viscosity is much longer than the 
last stages of inspiral considered here.


\begin{figure}
\begin{center}
\includegraphics[angle=0,width=13cm]{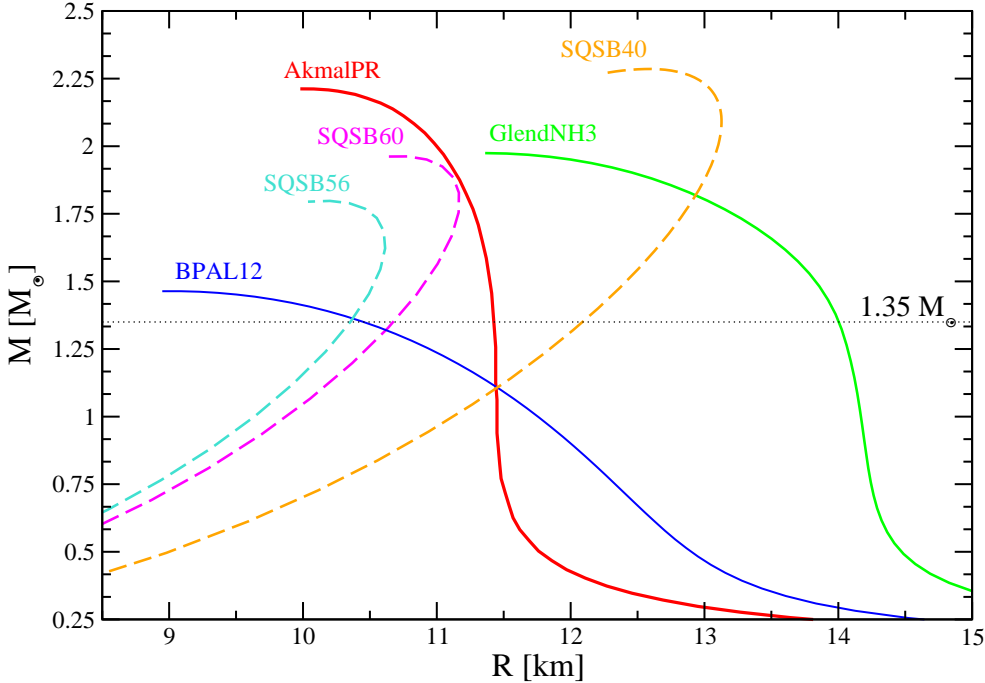}

\caption{Gravitational mass $M$ versus stellar radius $R$ for
sequences of static neutron stars described by three different nuclear
equations of state (solid lines) and  strange quark stars described
by different MIT bag models (dashed lines).
}  
\label{fig1}                                 
\end{center}                                 
\end{figure}

Strange stars are currently considered as a possible alternative to
neutron stars as compact objects (see e.g. \cite{Weber04, Madsen99,
Gondek03} and references therein). Typically, they are modeled with an
EOS based on the MIT-bag model in which quark confinement is described
by an energy term proportional to the volume (e.g. \cite{AlcocFO86,
HaensZS86}) in which quark confinement is described by an energy term
proportional to the volume \cite{Fahri84}. There are three physical
quantities entering the MIT-bag model: the mass of the strange quarks,
$m_{\rm s}$, the bag constant, $B$, and the strength of the QCD
coupling constant $\alpha$. In the framework of this model the quark
matter is composed of massless u, d quarks, massive s quarks and
electrons.  Strange stars are self-bound objects, having high density
(in the range $\sim 3 - 6.4\ [10^{14}{\rm g/cm}^3] $ \cite{Gondek03})
at the surface. Dashed lines in Fig. 1 correspond to sequences of
static strange quark stars described by three different sets of
parameters of the MIT-bag model:
{\bf SQS56} - the standard MIT bag model: $m_{\rm s}c^2=200\ {\rm MeV}$,
$\alpha=0.2$, $B=56~{\rm MeV/fm^3}$;
{\bf SQSB60} -  the simplified MIT bag model with  $m_{\rm s}=0$, $\alpha=0$;
$B=60\ {\rm MeV/fm^3}$;
{\bf SQSB40} - the ''extreme'' MIT bag model (relatively low strange 
quark mass and $B$ but high $\alpha$) : $m_{\rm s}c^2=100\ {\rm MeV}$,
$\alpha=0.6$, $B=40\ {\rm MeV/fm^3}$.

\begin{figure}
\begin{center}
\includegraphics[angle=0,width=13cm]{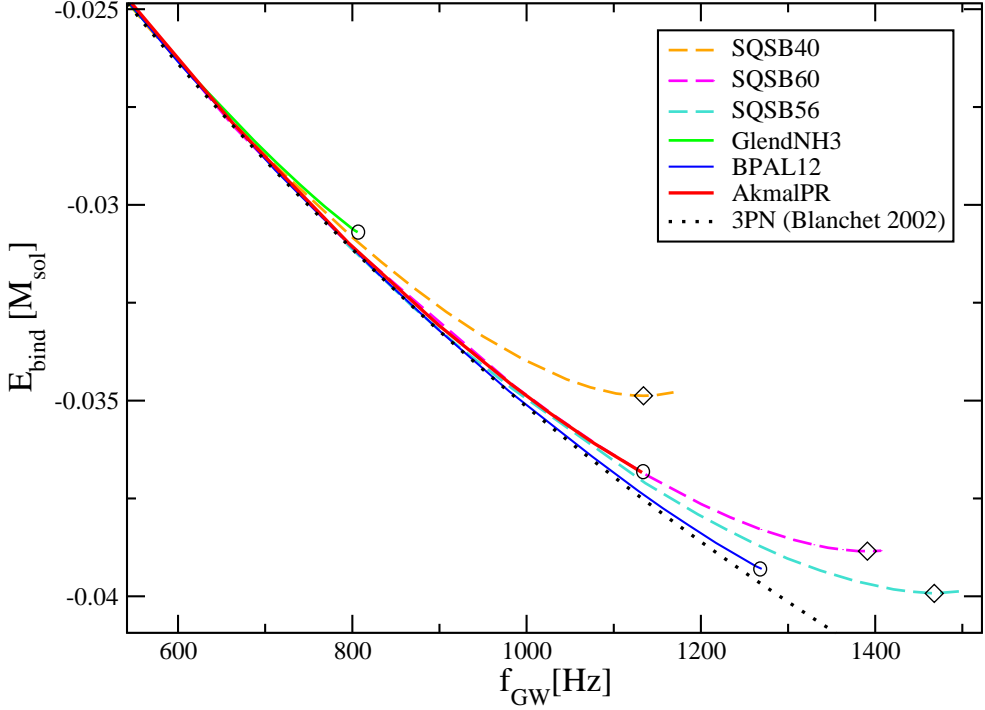}

\caption{Binding energy as a function of gravitational wave frequency
(twice the orbital frequency) along evolutionary sequences of
irrotational binaries. The solid lines denote neutron stars, the
dashed one strange quark stars and the dotted line point-mass binaries
in the third order Post-Newtonian approximation \cite{Blanc02}. Each
evolutionary sequence finishes at the innermost stable circular orbit.
The diamonds correspond to dynamical orbital instability while circles
to the mass-shedding limit.}
\label{fig2}                                 
\end{center}                                 
\end{figure}

\section{Numerical method}
In the inspiral phase, since the timescale of orbital shrinking
due to the emission of gravitational waves is longer than the orbital
period, one may consider a binary system to be in a quasi-equilibrium
state (helical Killing vector approximation).  For each EOS, we
construct so called an {\em evolutionary sequence} by calculating a
sequence of quasi-equilibrium configurations with fixed baryon mass
and decreasing orbital separation. We assume that the  timescale in which
 the neutron star adjusts to the current gravitational potential
is much shorter that the timescale of the orbital
shrinking. The present computations of close binary neutron star or
strange star systems rely on the assumption of irrotational flow of
the fluid and a conformally flat spatial 3-metric
(Isenberg-Wilson-Mathews approximation).  In order to calculate the
last orbits of inspiral phase of binary compact stars we use a
numerical code which solves the five coupled, nonlinear, elliptic
equations for the gravitational field, supplemented by an elliptic
equation for the velocity potential of irrotational flows
(see\cite{LimouGG05} for a discussion on different boundary conditions
in the case of strange stars and neutron stars).  The code has been
already used successfully for calculating the final phase of the
inspiral of compact stars
\cite{TanigGB01,TanigG02a,TanigG03,Bejger04,LimouGG05}.  This code is
built upon the C++ library {\sc Lorene} ({\tt
http://www.lorene.obspm.fr}).  The complete description of the
resulting general relativistic equations, the whole algorithm, as well
numerous tests of the code can be found in \cite{BNSmet01}. Additional
tests have been presented in Sect.~III of \cite{TanigG03}.

\section{Results}
In Fig. 2 we show the evolution of equal mass binary neutron stars
(solid lines) and strange stars (dashed lines) having total
gravitational mass $2.7 M_\odot$ at infinity. The binding energy
$E_{\rm bind}$ is defined as the difference between $M_{ADM}$
(Arnowitt-Deser-Misner mass - the total mass-energy of a
binary system  \cite{TanigG03}) and the total mass of the system at
infinite separation. This energy is equal to total energy emitted by a binary
system in gravitational waves. The frequency of gravitational waves is
twice the orbital frequency. Comparison of our numerical results with
3rd order PN point masses calculations \cite{Blanc02} reveals a good
agreement for small frequencies (large separations). The deviation
from PN curves at higher frequencies (smaller separation) is due to
hydrodynamical effects, which are not taken into account in the PN
approach.  A turning point of $E_{\rm bind}$ along an irrotational
evolutionary sequence indicates the orbital (dynamical) instability
\cite{FriedUS02}.  This instability originates both from relativistic
effects and hydrodynamical effects.  In the case where no turning
point of $E_{\rm bind}$ occurs along the sequence, the mass-shedding
limit (Roche lobe overflow) marks the end of the inspiral phase of the
binary system, since recent dynamical calculations for $\gamma = 2$
polytrope have shown that the time to coalescence was shorter than one
orbital period for configurations at the mass-shedding limit (i.e. see
\cite{MarronDSB04}).  Thus the physical inspiral of binary compact
stars terminates by either the orbital instability (turning point of
$E_{\rm bind}$) or the mass-shedding limit. In both cases, this
defines the {\em innermost stable circular orbit (ISCO)}. The end of
inspiral phase strongly depends on EOS - for irrotational neutron star
binaries a quasi-equilibrium sequence terminates by mass-shedding
limit (circles at the end of each line in Fig 2) and for strange stars
by orbital instability (shown as diamonds in Fig. 2). The differences
in the evolution of binary strange stars and neutron stars stem from
the fact that strange stars are principally bound by additional force
than gravitation: the strong interaction between quarks. They are
self-bound objects having very high adiabatic index at the stellar
surface (see \cite{LimouGG05}).  Although the crust of a $1.35\,
M_\odot$ neutron star contains only a few percent of the stellar mass,
this region is easily deformed under the action of the tidal forces
resulting from the gravitational field produced by the companion
star. The end of inspiral phase of binary stars strongly depends on
the stiffness of matter in this region.  The frequency of
gravitational waves at the ISCO is one of potentially observable
parameters by the gravitational wave detectors. In addition so called
``break frequency'', a characteristic frequency where the power
emitted in gravitational waves decreases measurably
\cite{Faber02,Hughes02}, could be an observable quantity. {\it The
frequencies of gravitational waves at the depature point from the 3PN
aproximation and at the ISCO strongly depend on the EOS.} For
irrotational equal mass (of 1.35 $M_{\odot}$ at infinite separation)
binaries the ISCO frequency is $\sim 1100-1460\ {\rm Hz}$ for strange
stars described by the MIT bag model and between $ 800\ {\rm Hz}$ and
$1230\ {\rm Hz}$ for neutron stars described by nuclear EOS (see also
\cite{Oechslin04, Bejger04}). The 3rd PN approximations for point
masses derived by different authors are giving ISCO at very high
frequencies of gravitational waves $> 2\ {\rm kHz}$ (\cite{Blanc02,
DamourJS, DamouGG02}).

It should be mentioned that the frequency at ISCO found
in quasi-equilibrium approximation could differ from that obtained by the
numerical integration of the full set of time dependent Einstein equations.
The only comparison (according to the authors' knowledge) between these two
approches for binary neutron stars can be found in  Marronetti et
al. (2004) who showed that for a polytropic EOS with $\gamma=2$ and $M/R=0.14$
the frequency at the ISCO
at quasi-equilibrium approximation is $\sim 15\%$ higher than the
dynamical ISCO.
Thus full relativistic hydrodynamical calculations
are required in order to obtain detailed predictions for 
particular forms of the nuclear matter equations of state.

\begin{ack}                               
This work was partially supported by the ``Ayudas para movilidad de
Profesores de Universidad e Invesigadores espanoles y extranjeros''
program of the Spanish MEC, by the grants 1 P03D 005 30,
PBZ-KBN-054/P03/2001 and 5 P03D 01721, by the ``Bourses de recherche
2004 de la Ville de Paris'' and by the Associated European Laboratory
Astro-PF (Astrophysics Poland-France). DGR is grateful to the COSPAR
meeting organizers for support. MB was partially supported by the
Marie Curie Intra-european Fellowship MEIF-CT-2005-023644.
\end{ack}
\newcommand{\apjl}{ApJ}
\newcommand{\apj}{ApJ}\newcommand{\mnras}{MNRAS}
\bibliographystyle{plain}        
\bibliography{}           
{}                  
\end{document}